\documentclass[a4paper,11pt]{article}
\pdfoutput=1 

\usepackage{jinstpub} 
\usepackage[normalem]{ ulem }
\usepackage{soul}
\usepackage{xspace}
\usepackage{textcomp}
\usepackage{amsmath}

\usepackage{lineno}
\usepackage{verbatim}
\usepackage{booktabs}

\title{Commissioning and Operation of the Readout System for the SoLid Neutrino Detector}

\title{Commissioning and Operation of the Readout System for the SoLid Neutrino Detector}

\collaboration{The SoLid Collaboration}
\author[a]{Y.~Abreu,}
\author[i]{Y.~Amhis,}
\author[d]{G.~Ban,}
\author[a]{W.~Beaumont,}
\author[o]{S.~Binet,}
\author[i]{M.~Bongrand,}
\author[i]{D.~Boursette,}
\author[j]{B.~C.~Castle,}
\author[o]{H.~Chanal,}
\author[b]{K.~Clark,}
\author[k]{B.~Coup\'e,}
\author[o]{P.~Crochet,}
\author[b]{D.~Cussans,}
\author[a,e]{A.~De Roeck,}
\author[d]{D.~Durand,}
\author[h]{M.~Fallot,}
\author[k]{L.~Ghys,}
\author[h]{L.~Giot,}
\author[g]{K.~Graves,}
\author[d]{B.~Guillon,}
\author[h]{D.~Henaff,}
\author[g]{B.~Hosseini,}
\author[g]{S.~Ihantola,}
\author[i]{S.~Jenzer,}
\author[k]{S.~Kalcheva,}
\author[c]{L.~N.~Kalousis,}
\author[f]{M.~Labare,}
\author[d]{G.~Lehaut,}
\author[b]{S.~Manley,}
\author[i]{L.~Manzanillas,}
\author[k]{J.~Mermans,}
\author[f]{I.~Michiels,}
\author[o]{S.~Monteil,}
\author[f,k]{C.~Moortgat,}
\author[b,m]{D.~Newbold,}
\author[l]{J.~Park,}
\author[d]{V.~Pestel,}
\author[b]{K.~Petridis,}
\author[a]{I.~Pi~nera,}
\author[k]{L.~Popescu,}
\author[f]{D.~Ryckbosch,}
\author[j]{N.~Ryder,}
\author[g]{D.~Saunders,}
\author[i]{M.-H.~Schune,}
\author[h]{M.~Settimo,}
\author[i,n]{L.~Simard,}
\author[g]{A.~Vacheret,}
\author[f]{G.~Vandierendonck,}
\author[k]{S.~Van Dyck,}
\author[c]{P.~Van Mulders,}
\author[a]{N.~van Remortel,}
\author[a,c]{S.~Vercaemer,}
\author[a]{M.~Verstraeten,}
\author[h]{B.~Viaud,}
\author[j,m]{A.~Weber,}
\author[h]{F.~Yermia}
\affiliation[a]{Universiteit Antwerpen, Antwerpen, Belgium}
\affiliation[b]{University of Bristol, Bristol, UK}
\affiliation[c]{Vrije Universiteit Brussel, Brussel, Belgium}
\affiliation[d]{Normandie Univ, ENSICAEN, UNICAEN, CNRS/IN2P3, LPC Caen, 14000 Caen, France}
\affiliation[e]{CERN, 1211 Geneva 23, Switzerland}
\affiliation[f]{Universiteit Gent, Gent, Belgium}
\affiliation[g]{Imperial College London, Department of Physics, London, United Kingdom}
\affiliation[h]{Universit\'e de Nantes, IMT Atlantique, CNRS, Subatech, France}
\affiliation[i]{LAL, Univ Paris-Sud, CNRS/IN2P3, Universit\'e Paris-Saclay, Orsay, France}
\affiliation[j]{University of Oxford, Oxford, UK}
\affiliation[k]{SCK-CEN, Belgian Nuclear Research Centre, Mol, Belgium}
\affiliation[l]{Center for Neutrino Physics, Virginia Tech, Blacksburg, Virginia, 24061, USA}
\affiliation[m]{STFC, Rutherford Appleton Laboratory, Harwell Oxford, United Kingdom}
\affiliation[n]{Institut Universitaire de France, F-75005 Paris, France}
\affiliation[o]{Universit\'e Clermont Auvergne, CNRS/IN2P3, LPC, F-63000 Clermont-Ferrand, France.}
\emailAdd{d.saunders@imperial.ac.uk, giel.vandierendonck@ugent.be}

\abstract{
    The SoLid experiment aims to measure neutrino oscillation at a baseline of 6.4\,m from the BR2 nuclear reactor in Belgium. Anti-neutrinos interact via inverse beta decay (IBD), resulting in a positron and neutron signal that are correlated in time and space. The detector operates in a surface building, with modest shielding, and relies on extremely efficient online rejection of backgrounds in order to identify these interactions. A novel detector design has been developed using 12800 5\,cm cubes for high segmentation. Each cube is formed of a sandwich of two scintillators, PVT and $^6$LiF:ZnS(Ag), allowing the detection and identification of positrons and neutrons respectively. The active volume of the detector is an array of cubes measuring 80x80x250\,cm (corresponding to a fiducial mass of 1.6\,T), which is read out in layers using two dimensional arrays of wavelength shifting fibres and silicon photomultipliers, for a total of 3200 readout channels. Signals are recorded with 14\,bit resolution, and at 40\,MHz sampling frequency, for a total raw data rate of over 2\,Tbit/s. In this paper, we describe a novel readout and trigger system built for the experiment, that satisfies requirements on: compactness, low power, high performance, and very low cost per channel. The system uses a combination of high price--performance FPGAs with a gigabit Ethernet based readout system, and its total power consumption is under 1\,kW. The use of zero suppression techniques, combined with pulse shape discrimination trigger algorithms to detect neutrons, results in an online data reduction factor of around 10000. The neutron trigger is combined with a large per-channel history time buffer, allowing for unbiased positron detection. The system was commissioned in late 2017, with successful physics data taking established in early 2018.
}

\keywords{
Neutrino detectors; Front-end electronics for detector readout; Analogue electronic circuits; Trigger detectors; Data acquisition circuits
}

\begin{document}
\maketitle
\flushbottom

\section{Introduction}
\subsection{The SoLid Experiment}
SoLid is designed to measure neutrino oscillations at very short baselines, $\mathcal{O}(10)$\,m, from a nuclear reactor core. Hints of reactor neutrino oscillations at this energy and distance scale arise from both the reactor and gallium anomalies~\cite{KOPP}. Measurements of the rate of $\overline{\nu_{e}}$ emitted by reactor cores show a deficit at $\sim 3 \sigma$ significance when compared with expectations. Deficits of a similar significance are also observed in measurements of the $\nu_{e}$ rate emitted by radioactive sources. One proposed explanation is the existence of an additional non-interacting `sterile' neutrino state, and a corresponding mass eigenstate; this fourth neutrino could influence the neutrino flavour transitions (oscillations) at very short distances. The existence of such oscillations can be tested using measurements of the $\overline{\nu_{e}}$ energy spectrum as a function of distance from a neutrino source.

\begin{figure}
\begin{center}
\includegraphics[width=0.6\textwidth]{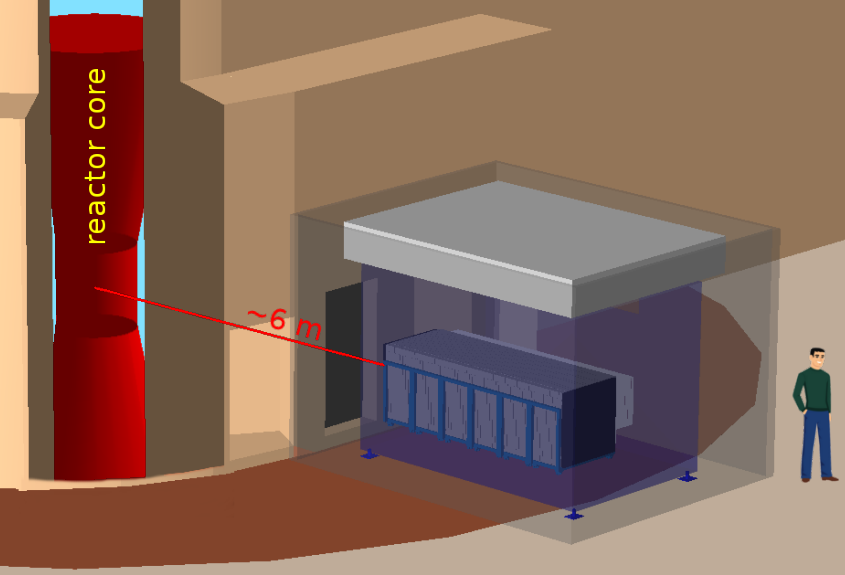}
\caption{Schematic showing the placement of the SoLid detector in the BR2 reactor containment building. The detector itself is placed in a customised shipping container, and surrounded by a 50\,cm wall of passive water shielding. A roof structure supports 50\,cm of High-density polyethylene (HDPE) passive shielding.}
\label{solid_at_br2}
\end{center}
\end{figure}

The BR2 reactor core is especially suitable given its small core diameter of 0.5\,m (see figure~\ref{solid_at_br2}). The fuel composition is predominantly ($>$ 90$\%$) $^{235}$U, which is of particular interest to the nuclear physics community, who await updated $\overline{\nu_{e}}$ energy spectrum measurements from highly enriched fuels to resolve the 5\,MeV distortion observed by previous reactor experiments~\cite{fiveMeVbump}. The reactor is powered to around 60\,MW for typically half the year in evenly spread 1 month cycles. The space in the reactor hall is sufficient for a relatively compact detector to be placed with modest passive shielding.

Compared to previous neutrino oscillation experiments that operate at longer baselines, this environment is particularly challenging. The detector has to be placed on the surface, with negligible overburden to shield from cosmic-ray backgrounds. Additionally, the reactor itself produces a large rate of gamma rays during operation that can further contribute to backgrounds. Previous experience of running a 288\,kg (20$\%$ scale) prototype of the detector in spring 2015~\cite{sm1}, demonstrated the need for a controlled temperature environment to reduce and stabilise the silicon photomultiplier (SiPM) dark count rate, as well as a need for electromagnetic shielding to prevent pickup of electronic noise. The detector is scheduled to run for around three years. Efficient online signal tagging is required in order to reach the experiment's physics aims in this time period. 

\subsection{Detector Design}
Anti-neutrinos are detected in SoLid via inverse beta decay (IBD), resulting in a positron and neutron signal that are correlated in space and time. To take advantage of this spatial correlation, the SoLid detector is highly segmented, with its detection volume formed of 12800 5\,cm cubes. This corresponds to the scale of the mean separation between the positron and neutron interaction. The bulk of each cube is polyvinyltoluene (PVT) based scintillator that offers high light output and a linear energy response. Sheets of $^6$LiF:ZnS(Ag) are placed on two faces of each cube to facilitate neutron detection. In order for each cube to be optically isolated, it is wrapped in white Tyvek \textregistered. Neutrons may be captured on the lithium via the interaction: 

\begin{equation}
    n + \mathrm{^{6}Li} \rightarrow \alpha + \mathrm{^{3}_{1}H}\ (4.8 \text{ MeV})
\end{equation}

The alpha and tritium particles deposit energy in the ZnS(Ag) causing scintillation. These heavy nuclei scintillations are referred to as \textit{nuclear signals} (NS). Crucially, the scintillation timescale of nuclear signals is considerably slower at $\mathcal{O}(1)$\,\textmu{}s, than the PVT scintillation at $\mathcal{O}(1)$\,ns. Nuclear signals are characterised by a set of sporadic pulses emitted over several microseconds. Pulse shape discrimination (PSD) techniques can be used to identify the nuclear signals with high efficiency and purity. These are used both in offline data analysis and, in simplified form, in the trigger. The full signal waveforms are therefore required for offline analysis. A sampling speed of around 40\,MHz is sufficiently fast in order to perform effective PSD, whilst providing adequate time resolution. This has been demonstrated with smaller scale lab setups and by a large scale prototype of the full detector~\cite{sm1}.

In IBD interactions, the positron is detected immediately via scintillation in PVT, and the neutron is detected after thermalisation and capture. The separation of the positron and neutron is 2 cubes or less in 90\% of IBD interactions. The gamma rays resulting from the annihilation of the positron travel up to 30\,cm, and can deposit energy in other neighbouring cubes. The mean time interval between the positron and neutron scintillation signals is around 60\,\textmu{}s, and the neutron capture efficiency of this configuration is around 66\%. For a reactor power of 60\,MW, the expected rate of neutrino captures in the detector is approximately 1200 per day (around 1 per 100 seconds). 

The cubes are arranged in 50 planes of 16$\times$16 cubes. Light from each detector plane is read out via a 2D array of vertical and horizontal wavelength shifting fibres that sit in grooves machined in the PVT. Two fibres sit along each row and column of cubes, giving 64 fibres per plane. The use of two fibres per vertical/horizontal direction enhances the overall light collection efficiency, as well as providing channel redundancy. One end of each fibre is coupled to a second-generation Hamamatsu type S12572-050P SiPM, whilst on the other end there is a mirror. The light yield has been measured to be 15 pixel avalanches (PA) per fibre per MeV of deposited energy, for a typical operational SiPM over-voltage of 1.5\,V. At this over-voltage, a dynamic range of $\mathcal{O}(10^4)$ ADC counts is required (i.e 14 bit resolution). The resolution is on the one hand driven by the need for adequate resolution for small signals to cleanly separate the 1\,PA peak from background and on the other hand by the requirement for linear response to large signals such as cosmic ray muons.

The entire detector is placed in a light-tight and thermally insulated shipping container. The inside of the container is cooled to 10$^{\circ}$C, which reduces the dark count rate of the SiPMs, whilst also providing a thermally controlled environment.

\subsection{Readout System Requirements}
The physics program described above results in the following requirements for the SoLid readout system:

\subsubsection{Functional Requirements}

\begin{itemize}
    \item Efficient online tagging of the anti-neutrino signal, with effective discrimination of signal from background
    \item Digitisation, continuous capture, and buffering of streaming data from 3200 channels of optical readout, synchronously across the detector
    \item Flexible event-building and transmission of data to storage
\end{itemize}

\subsubsection{Performance Requirements}
\begin{itemize}
    \item Digitisation of analogue signals at 40\,Msample/s (25\,ns per sample)
    \item Synchronisation of channels to within 1\,ns 
    \item Selection of IBD candidates with $>$70\% trigger efficiency and data reduction factor $\mathcal{O}(10^4)$
    \item Up to 1\,Gb/s data recording to permanent storage
    \item Readout of selectable space and time regions around the IBD candidate
    \item Negligible dead time
\end{itemize}

\subsubsection{Technical Requirements}
\begin{itemize}
    \item Modular design driven by integration requirements
    \item Shielding from external pickup noise
    \item Low cost compared to commercially available systems ($<$\$75 per channel)
    \item Low power consumption ($<$1\,kW total, or $<$0.3\,W per channel)
    \item High reliability for unattended operation and restricted access
\end{itemize}

This paper describes a readout system for the SoLid experiment that fulfils all these requirements. Section~\ref{sec:design} describes the hardware design of the system. First results and characterisation of the detector channels are presented in section~\ref{sec:channel_characterisation}. Section~\ref{sec:trigger} describes the strategy and implementation of online triggers and data reduction techniques, including their optimisation. Finally, section~\ref{sec:performance} gives in situ performance results of the system as a whole, including the life-time of the detector as a function of trigger rate, and stability over time.

\section{Readout System Design}
\label{sec:design}
In order to achieve the above requirements, particularly for mechanical flexibility at a low cost, most components of the readout system are of custom design. This allows the system to be integrated with the detector directly inside the chilled container, removing the need for long analogue cabling, which both reduces the risk of electromagnetic interference and improves the ease of installation. The system modularity is split into three levels: a single detector plane, a set of ten detector planes (i.e a detector module), and multiple modules (i.e the full detector). Partitioned running of the system in each of these modularities is supported. Each level is now discussed in turn. 

\subsection {Detector Planes}
\begin{figure}
\begin{center}
\includegraphics[width=0.6\textwidth]{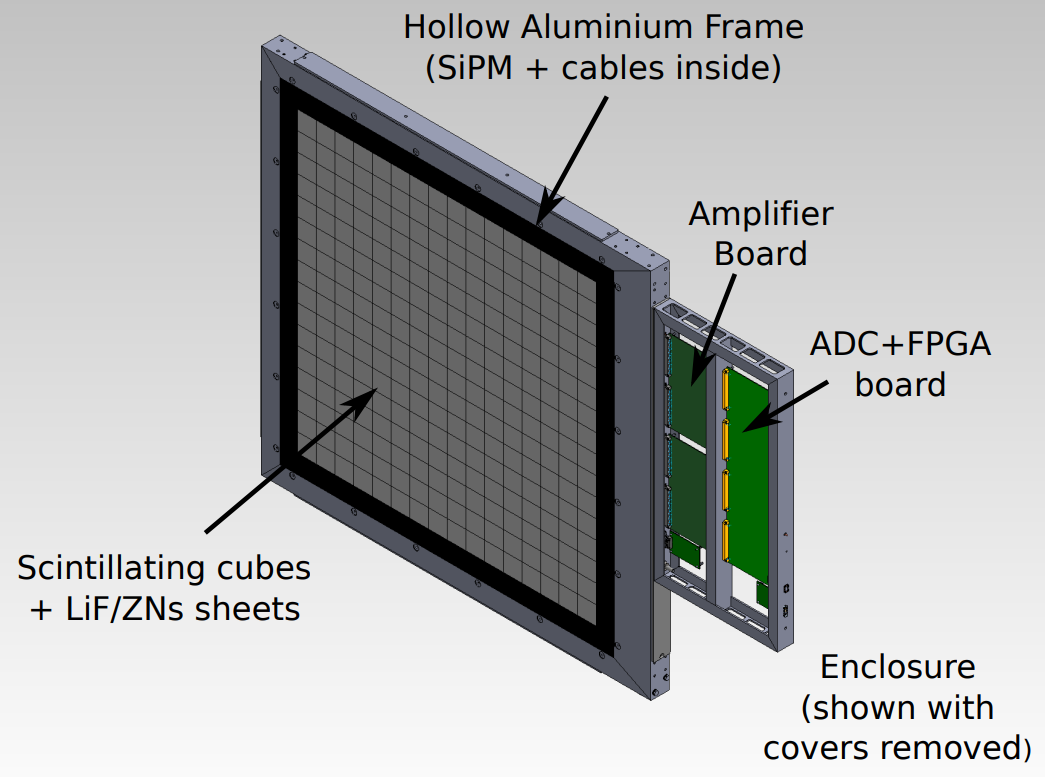}
\caption{Diagram of a single detector plane with a mounted electronics enclosure.}
\label{fig_plane_digram}
\end{center}
\end{figure}

For each detector plane, the 64 SiPMs are placed in a hollow aluminium frame. One side of the frame has an interface plate with connectors for the SiPMs. The SiPMs connect to this plate using twisted-pair ribbon cables that terminate into insulation displacement connectors (IDC). The other side of the interface plate connects to an aluminium electronics enclosure, which is mounted directly on the frame side, and houses all required electronics to run a single detector plane (see figure~\ref{fig_plane_digram}). Each enclosure is electrically connected to its frame, and so each detector plane acts as a Faraday cage, providing shielding from outside electronics noise contributions. The enclosures can be quickly attached and detached from each frame, allowing for ease of integration and replacement. 

\begin{figure}
\begin{center}
\includegraphics[width=0.6\textwidth]{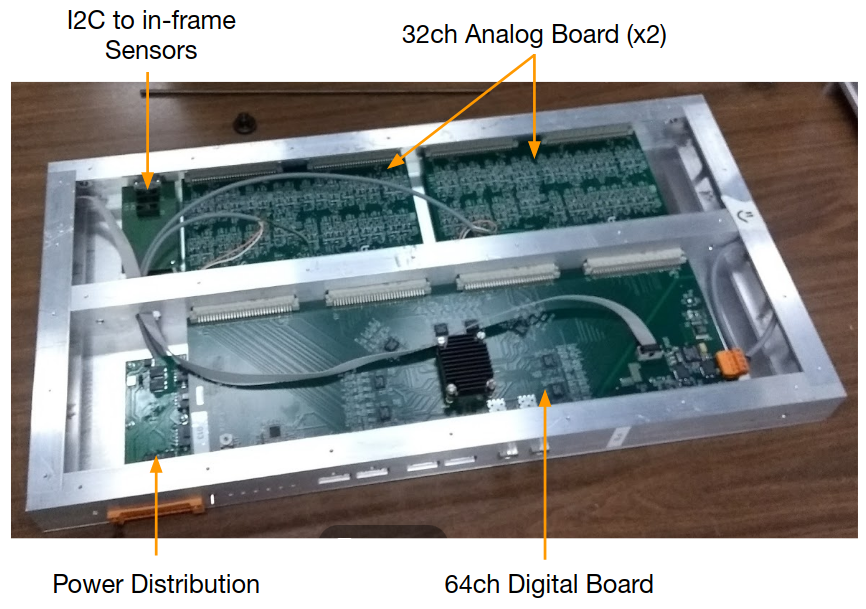}
\caption{Picture of a fully assembled electronics enclosure for one detector plane.}
\label{fig_enclosure_pic}
\end{center}
\end{figure}

Each enclosure contains two \emph{analogue boards}, each containing amplification and biasing circuitry, and a single \emph{digital board} (see below). Additionally, there are two smaller boards: one for I$^2$C communication with four in-frame environmental sensors, and another board for power distribution. A picture of a complete assembled electronics enclosure is shown in figure~\ref{fig_enclosure_pic}. The top and bottom sides of the enclosure have openings to allow air flow to cool the electronics.

\subsubsection{Analogue Electronics}
\begin{figure}
\begin{center}
\includegraphics[width=0.6\textwidth]{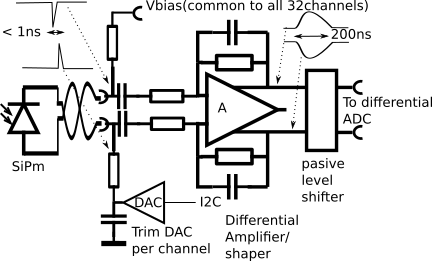}
\caption{A schematic block diagram of the analogue electronics in the SoLid detector.}
\label{fig_analogueElect}
\end{center}
\end{figure}

The 64 frame SiPMs connect to two 32-channel analogue boards. A schematic block diagram of the analogue electronics is shown in figure~\ref{fig_analogueElect}. The cathodes of all SiPMs connected to one board are biased from a common 70\,V supply. The breakdown voltage and corresponding gain variations of the individual SiPMs are equalised by applying individual trim bias voltages (0-4\,V) to each SiPM. The setting of SiPM over-voltage is a compromise between photon detection efficiency, pixel cross-talk and thermal dark count rate. This can have a significant impact on the neutron detection efficiency, given the low amplitude and sporadic nature of nuclear signals. Experience from prototype setups suggests values between 1.5\,V and 1.8\,V are suitable, and this is optimised later for neutron efficiency. 

Since the SiPM pulses are very fast (a few ns) compared to the digital sampling (25\,ns), the pulses are shaped to slow the signal over several samples. Signal time estimation is performed on the ADC-sampled waveform, rather than by a discrimination technique on the full analogue signal. The effective time resolution is therefore a function of the ADC sampling period, the uncertainty on the sample clock phase, and the shaping time of the analogue electronics. By using a shaping time corresponding to several ADC samples, and a constrained fit to the pulse shape, a more accurate timestamp and amplitude can be measured offline.

Due to the AC coupling between the SiPMs and the analogue board amplifier, the mean signal over long time periods is forced to be zero, and as a consequence, the pedestal of each channel depends on that channel's signal rate. In practice, it was observed that large changes ($>$0.5\,V) to the bias voltage, which can change the SiPM dark count rate, can cause significant changes in the pedestal, requiring the pedestals to be re-measured for input to the trigger. However, other factors that affect the signal rate, including the presence of intense calibration sources, or the reactor on-off transition, resulted in less significant changes, and do not require a re-calibration. 

\subsubsection{Digital Electronics}
The two analogue boards connect directly to a 64-channel digital board for digitisation and online data processing. Each digital board has eight 8-channel 14 bit 40\,Msample/s ADCs, and is controlled and read out over a 1\,Gb/s optical Ethernet connection. A Phase-Locked Loop (PLL) is included because the data transfer from ADC to FPGA takes place at a multiple of the sample clock frequency. The use of a high-performance crystal-locked PLL also allows the reduction of sampling clock jitter and provides the possibility to operate the board with an internally generated clock, or run synchronised to an external clock signal. Two duplex 2.5\,Gb/s links carried on copper cables allow data (e.g trigger signals) from each digital board to be propagated to all other detector planes along a daisy chain. An on-board over-temperature shutdown is included. JTAG connectors are included for remote firmware programming.

The trigger and readout logic is implemented in a Xilinx Artix-7 FPGA device (XC7A200). The sizing of the FPGA device is driven by the buffer RAM requirement, which is in turn dictated by the trigger rate and space-time region read out, convoluted with the waveform compression factor due to zero suppression. This device offers sufficient block RAM resources for likely running conditions at a modest cost, and was conveniently available in commercial sub-assemblies~\cite{trenz}.

\subsection{Detector Modules}
\begin{figure}
\begin{center}
\includegraphics[width=0.6\textwidth]{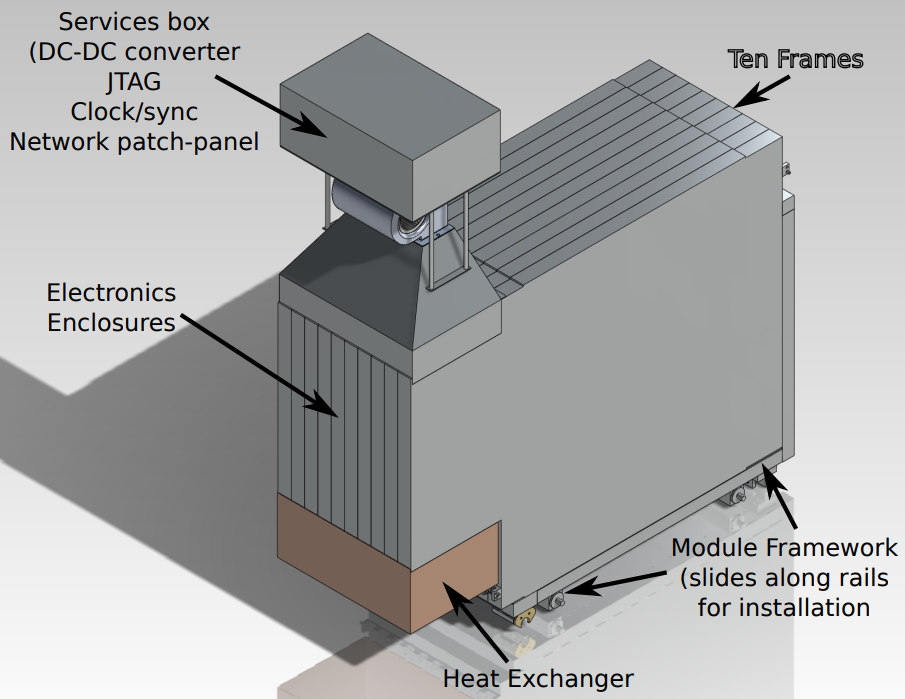}
\caption{Diagram of a detector module, composed of ten detector planes, a module services box and a heat exchanger.}
\label{fig_module_diagram}
\end{center}
\end{figure}

Ten planes are grouped together to form a mechanically independent detector module (see figure~\ref{fig_module_diagram}). A module clock-board, which can operate in both master and slave mode, provides a common clock fan-out, synchronising the ten associate digital boards. The clock-board is placed in a module services box that is mounted above the ten electronics enclosures. The services box also include a DC-DC converter to power the module, as well as a Minnow single board computer to program the clock-board and detector plane FPGA firmware via a JTAG fan-out.

In order to cool the enclosures, fans are mounted between the services box and the plane electronics enclosures, pushing air downwards. A heat exchanger is placed below the enclosures, which removes heat generated by the electronics ($\sim$200\,W per module), and also acts as the cooling source for the enclosing container.

\subsection{Full Detector Configuration}
Finally, modules are grouped and operate synchronously once connected to an additional master clock-board, which drives the module slave clocks. The final SoLid detector configuration comprises of five detector modules.

\subsubsection{Cost Summary}
Table 1 shows the hardware costs of the readout system. The final cost of the system is around \$220k, or \$70 per channel, with 80\% of the budget dedicated to the per plane electronics.

\begin{figure}
\begin{center}
\includegraphics[width=0.6\textwidth]{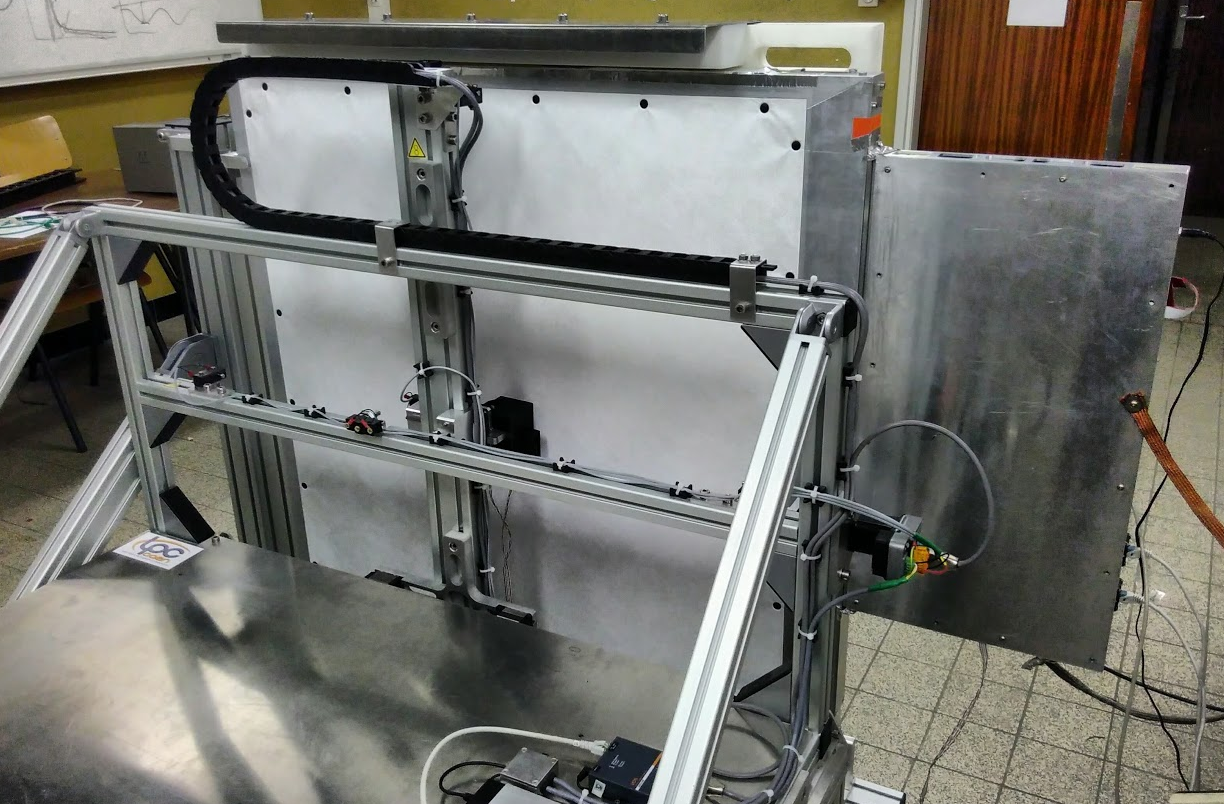}
\caption{The Calipso quality assurance setup, which used a full scale prototype SoLid electronics enclosure. The enclosure, dubbed `Zoidberg', read out over 100 million neutrons during the QA campaign to ensure a uniform plane response.}
\label{fig_pic_calipso}
\end{center}
\end{figure}

\begin{table}[]
\begin{center}
\small

\begin{tabular}{@{}lrrr@{}}
\toprule \midrule
Component & \multicolumn{1}{l}{\begin{tabular}[c]{@{}l@{}}Unit \\ Cost \\(\$)\end{tabular}} & \multicolumn{1}{l}{\begin{tabular}[c]{@{}l@{}}Number of \\ Channels\end{tabular}} & \multicolumn{1}{l}{\begin{tabular}[c]{@{}l@{}}Unit Cost \\ Per-Channel \\(\$)\end{tabular}} \\ \midrule 
\textbf{Per Detector Plane} & & & \\ 
Digital Board (pc051c) & 2000 & 64 & 31 \\
Analogue Board & 650 & 32 & 20 \\ 
FPGA (xc7a200) & 350 & 64 & 5.5 \\ 
Mechnical Enclosure & 250 & 64 & 4.4 \\ 
\textit{Plane Sub-Total} & \textit{3900} & \textit{64} & \textit{61} \\ \midrule
\textbf{Per Detector Module} & & & \\ 
Clock-board \& FPGA & 2000 & 640 & 3.0 \\ 
PSU PCB & 650 & 640 & 1.0 \\ 
JTAG Fanout & 520 & 640 & 0.8 \\ 
\begin{tabular}[c]{@{}l@{}}Other Service \\ Box Hardware\end{tabular} & 460 & 640 & 0.6 \\ 
\textit{Module Sub-Total} & \textit{3600} & \textit{640} & \textit{5.6} \\ \midrule
\textbf{Full Detector} & & & \\ 
Network Switch (x3) & 2000 & 3200 & 0.6 \\  \midrule
\textit{\textbf{Total Full Detector}} & \textit{\textbf{220k}} & \textit{\textbf{3200}} & \textit{\textbf{70}} \\  \midrule \bottomrule
\rule{0pt}{0.1ex} 

\end{tabular}
\label{tab_costs}
\caption{Costs of the SoLid readout system, split by modularity, and in descending cost per channel.}
\end{center}
\end{table}

\subsection{Prototyping}
The digital and analogue electronics underwent two stages of prototyping. A small 8-channel system served as the first prototype~\cite{twepp}, and a second prototyping round saw full-scale boards. Once electronically commissioned, a 64-channel setup was used for detector quality assurance (QA) tests of each 256 cube detector plane. In addition to testing the long term stability of the electronics, this QA setup served as a test platform for trigger development, data acquisition, and detector monitoring. Each cube of each detector plane was studied with several radioactive sources including $^{22}$Na and Cf, which were moved across the plane face using a dedicated calibration robot: Calipso \cite{qa_paper}. A picture of the setup is shown in figure~\ref{fig_pic_calipso}.

\subsection{Readout Software and Communication}
\begin{figure}
\begin{center}
\includegraphics[width=0.6\textwidth]{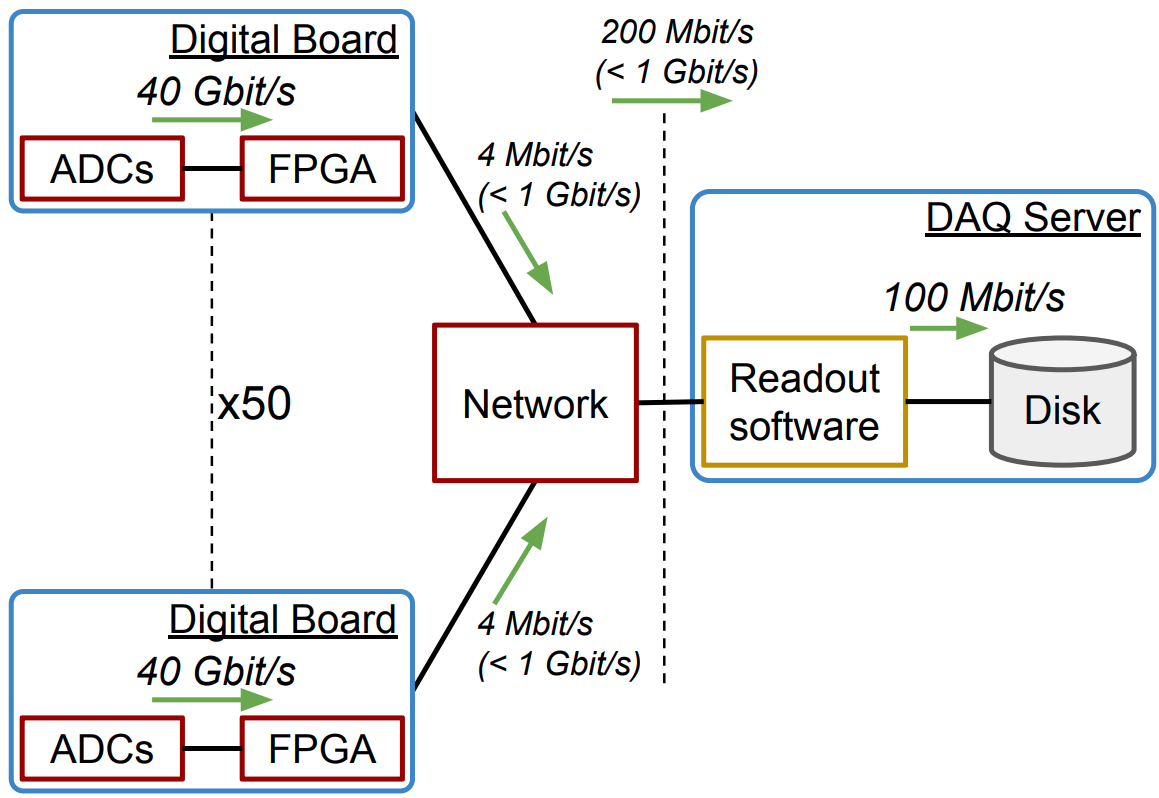}
\caption{SoLid data flow from ADCs to disk. The rates stated are typical for physics data taking mode. The upper limits are also marked: the network sets the upper limit for each detector plane, whereas the recorded data rate limit is due to the readout software.}
\label{fig_readout_flow}
\end{center}
\end{figure}

Figure~\ref{fig_readout_flow} shows the SoLid data flow, beginning with the digital board ADCs, through to data recorded on disk at the on-site DAQ server. Each digital board can be read out over a 1\,Gb/s optical ethernet port. During normal data taking operations, the data rate is considerably lower, at around 2\,Mb/s. During source runs, planes near the source can reach the maximum limit. The 2\,Tb/s total data rate out of the ADCs is reduced down to $\sim$200\,Mb/s recorded to disk.

The FPGAs are configured and read out using the IPbus protocol, which operates on top of UDP/IP and is designed for reliability and high performance~\cite{ipbus}. The readout software is written in the \emph{Golang} programming language, to take advantage of the language's intrinsic multi-threading and memory management tools. As a result, a Golang implementation of the IPbus software is now available: goIPbus. The software can configure all readout electronics components for data taking, and performs continuous retrieval of data from all fifty detector planes in parallel. A minimal parsing of the data is performed to check the format is as expected, and data from all planes are concatenated into a single data file, in the order of retrieval from the FPGAs. The readout software has been tested up to 1\,Gb/s - test results are shown in section~\ref{sec:DAQ_performance}. 

\section{Channel Characterisation}
\label{sec:channel_characterisation}

\subsection{ADC Alignment}
For the FPGA to be able to correctly receive the serial data streams from each ADC channel, the delay of each digital stream must be tuned so that the received data from all channels is aligned. Misalignments can be caused by the different placements of these devices on the large digital board, and are not expected to change significantly board-to-board. The alignment of each channel can be set using two parameters known as \textit{slip} and \textit{tap}. The slip parameter allows data to be delayed by an integer number of bits, and the tap parameter allows for finer increments. 

\begin{figure}
\begin{center}
\includegraphics[width=0.6\textwidth]{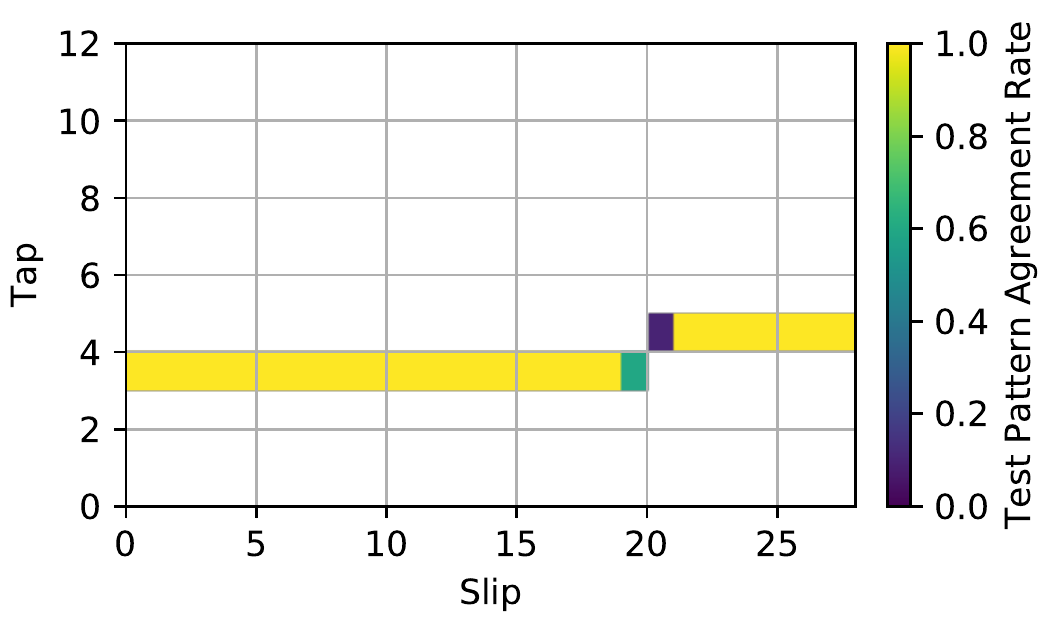}
\caption{ADC alignment scan results for channel 0. The colour axis shows the rate of agreement between input ADC test pattern and value read back.}
\label{fig_adc_alignment}
\end{center}
\end{figure}

The ADCs can be configured to return a fixed value (i.e a test pattern). Suitable alignment parameters can be found by comparing the data, which is read out via the FPGA, with the input test pattern. A typical example output from a 2D scan of tap and slip is shown in figure~\ref{fig_adc_alignment} for the first detector channel. There is a wide range of alignments that consistently return the input test pattern; the central value is chosen. This alignment procedure has been repeated for all detector channels. The spread in alignment parameters is small within both individual ADC chips, and all chips in the same position on the digital boards. A default set of alignments can be used for the majority of detector channels based only on the position of its ADC, although approximately 10\% of channels required specific per-channel alignments. 


\subsection{SiPM Characterisation}
\begin{figure*}
\begin{center}
\includegraphics[width=1.0\textwidth]{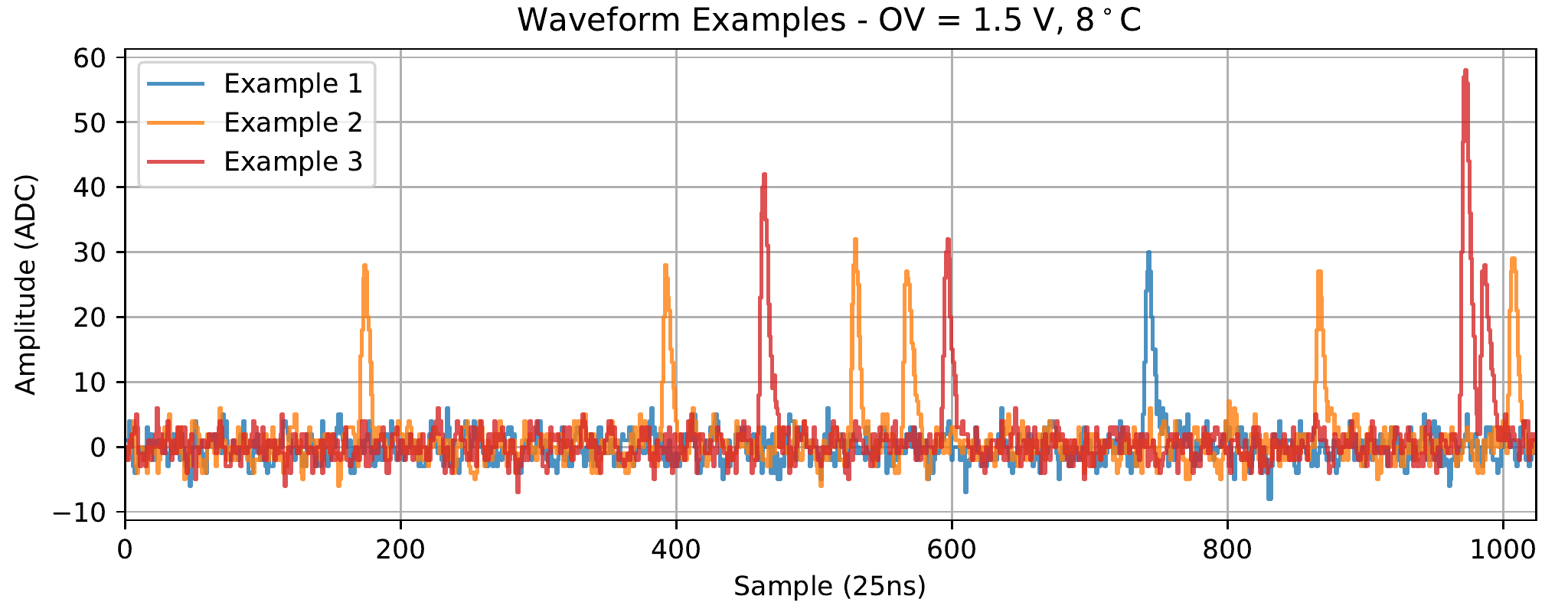}
\caption{Example waveforms from the first three detector channels (using random trigger and pedestal subtracted) of the complete SoLid detector.}
\label{fig_eg_waveforms}
\end{center}
\end{figure*}

\begin{figure*}
\begin{center}
\includegraphics[width=1.0\textwidth]{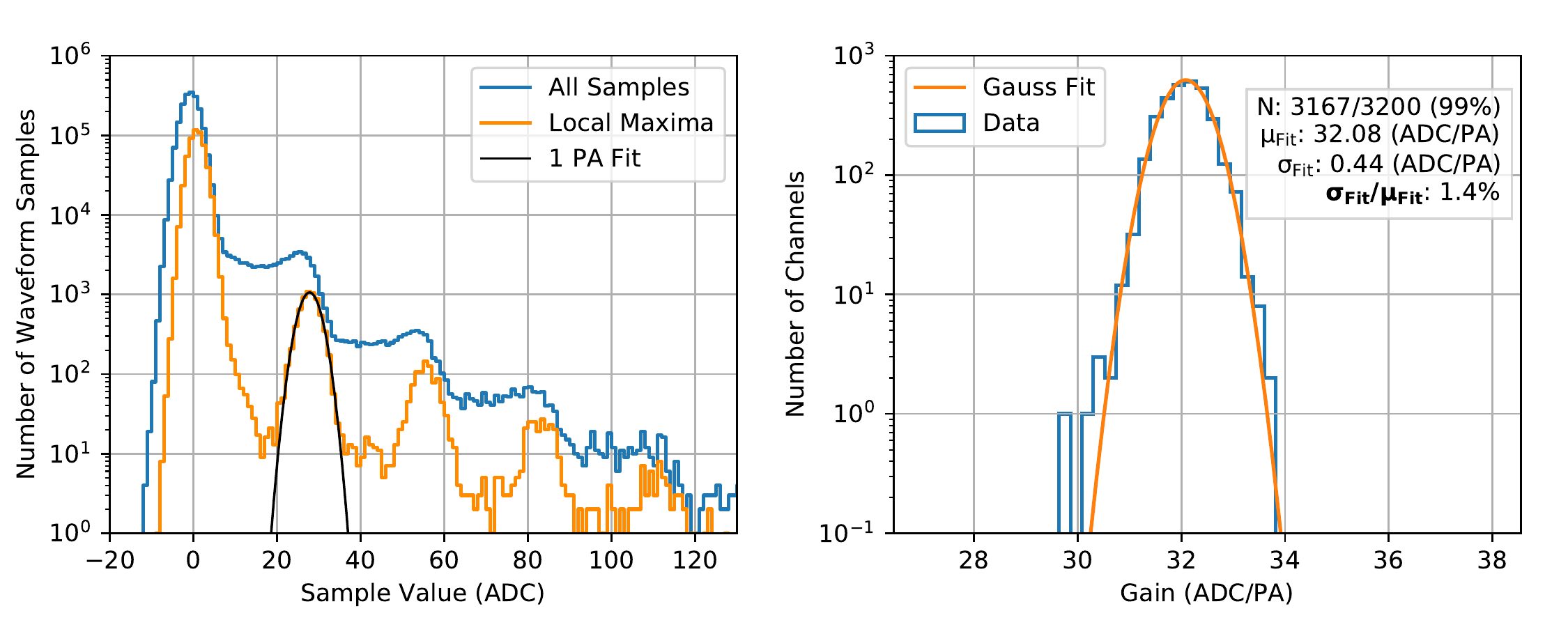}
\caption{Left: Spectrum of ADC samples for the first detector channel, with and without a local maxima filter applied (pedestal subtracted). The first pixel avalanche peak can be parameterised using a Gaussian curve, and the mean parameter is used as the channel gain measurement. Right: Spread of gain values across all operational SiPMs after the final iteration of the equalisation procedure.}
\label{fig_gain_results}
\end{center}
\end{figure*}

Example SiPM waveforms are shown in figure~\ref{fig_eg_waveforms}, and the distribution of all waveform samples from a single detector channel is shown in figure~\ref{fig_gain_results}. The quantisation of the pixel avalanche peaks is clearly visible, particularly when a local maximum filter is applied. The width of the pedestal is relatively small at around 2\,ADC counts rms, implying a low level of external noise pickup.

In addition to 15 channels masked due to difficulties in ADC alignment, 18 other channels are masked due to broken connections between the SiPMs and the readout electronics, and due to one damaged ADC on one of the digital boards. This results in a total of 33/3200 masked channels (1\%). There are no cases where a pair of SiPMs coupled to fibres along the same column/row of cubes are both masked, allowing all cubes to be read out with at least 3 active fibres.

\subsubsection{Gain Equalisation}
The SiPMs are equalised using the per-channel trim voltage to achieve a uniform single-photon amplitude response. The dominant source of non-uniformity is due to the large spread in SiPM breakdown voltage, where the standard deviation is around 2\,V for those in use in the detector (cf. an operational over-voltage of around 1.8\,V). Additional non-uniformities can be caused by variation in the differential gain with respect to over-voltage. This function is measured individually for every detector channel. The gain is taken as the position of the 1\,PA peak of the waveform amplitude distribution with a local maximum filter applied. This peak can be well fitted with a Gaussian curve, as demonstrated in figure~\ref{fig_gain_results}, and the mean of the Gaussian is taken as the gain value. The amplitude of the fit is constrained by the number of entries in the distribution, and the width is constrained by the typical fit width. These constraints result in a highly robust gain finding technique, with a failure rate of around 0.1\% in the range 1-3\,V over-voltage, and precision around 0.5\,ADC counts. 

\begin{figure}
\begin{center}
\includegraphics[width=.7\textwidth]{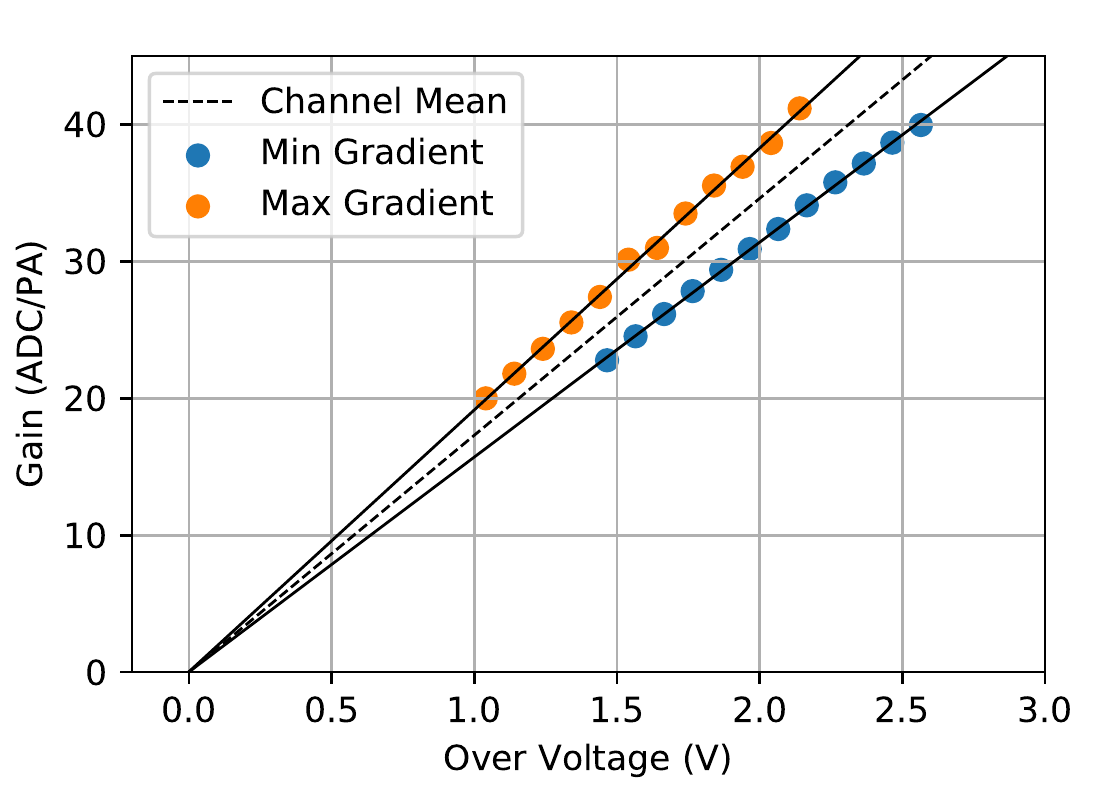}
\caption{Voltage scans for the two channels with extremal amplification rates (i.e min and max gradient). In the extreme cases, the amplification rate variation can cause a change in gain of around 10 \% at 1.8\,V over-voltage.}
\label{fig_v_scan}
\end{center}
\end{figure}

The variation in breakdown voltage and the over-voltage sensitivity necessitated an iterative voltage scanning procedure, with the appropriate parameters for a given SiPM channel successively refined between scans. Figure~\ref{fig_v_scan} shows the resulting mappings for the channels with extremal cases of gain differentials (i.e max and min gradient of gain v.s voltage). A linear fit is performed for each channel, and the output parameters stored in a database. Upon setting the bias voltage, the database is queried to look up the required bias voltage for a desired gain. Figure~\ref{fig_gain_results} shows that the spread in gain across channels after this procedure is 1.4\%, where the dominant uncertainty is the precision of the gain-finder itself. The desired gain is set to 32\,ADC counts per PA, which corresponds to a mean over-voltage of 1.8\,V. The spread increases to around 7\% when applying a bias voltage that depends only on the breakdown voltage (i.e a fixed over-voltage for all channels, without taking into account the per-channel gradient of the linear fit). The whole procedure is 100\% successful for all non-masked detector channels. The measured breakdown voltages are highly correlated with manufacturer values supplied by Hamamatsu, with a residual standard deviation of 0.1\,V.  

\subsubsection{Dark Count Rate}
\begin{figure}
\begin{center}
\includegraphics[width=0.6\textwidth]{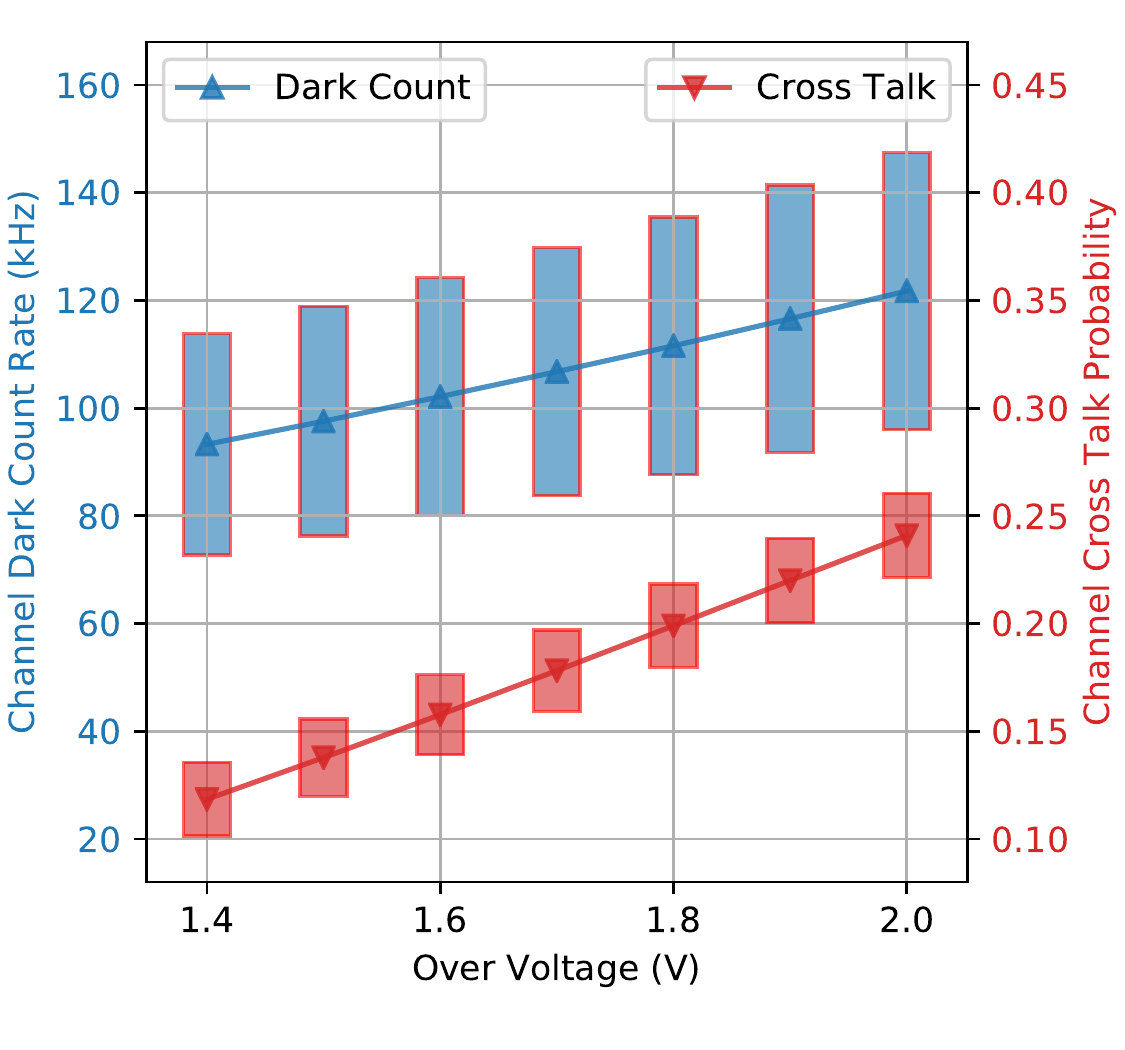}
\caption{Channel dark count rate and cross-talk as a function of over-voltage. The points show the channel mean, and the boxes show the standard deviation across all channels. The dark count rate is taken as the number of local maxima above 0.5\,PA. The cross-talk probability is taken as the ratio of the rates of the 2\,PA to 1\,PA signals for each channel.}
\label{fig_noise_metrics}
\end{center}
\end{figure}

The mean and spread of both the dark count rate and SiPM pixel cross-talk probability across detector channels is shown in figure~\ref{fig_noise_metrics} as a function of over-voltage. The mean dark count rate is 110\,kHz per channel at an over-voltage of 1.8\,V and at 11$^{\circ}$C, which is around a factor of three reduced compared with room temperature (20$^{\circ}$C), using the same over-voltage. The corresponding cross-talk probability is 20\%. These values are consistent with expectations from studies of SiPMs in environmentally controlled lab setups~\cite{wouter_thesis}. The spread in dark count rate is approximately uniform across the detector volume, with no apparent hot spots. 

\subsubsection{Zero Suppression Reduction Factor}
\begin{figure}
\begin{center}
\includegraphics[width=0.6\textwidth]{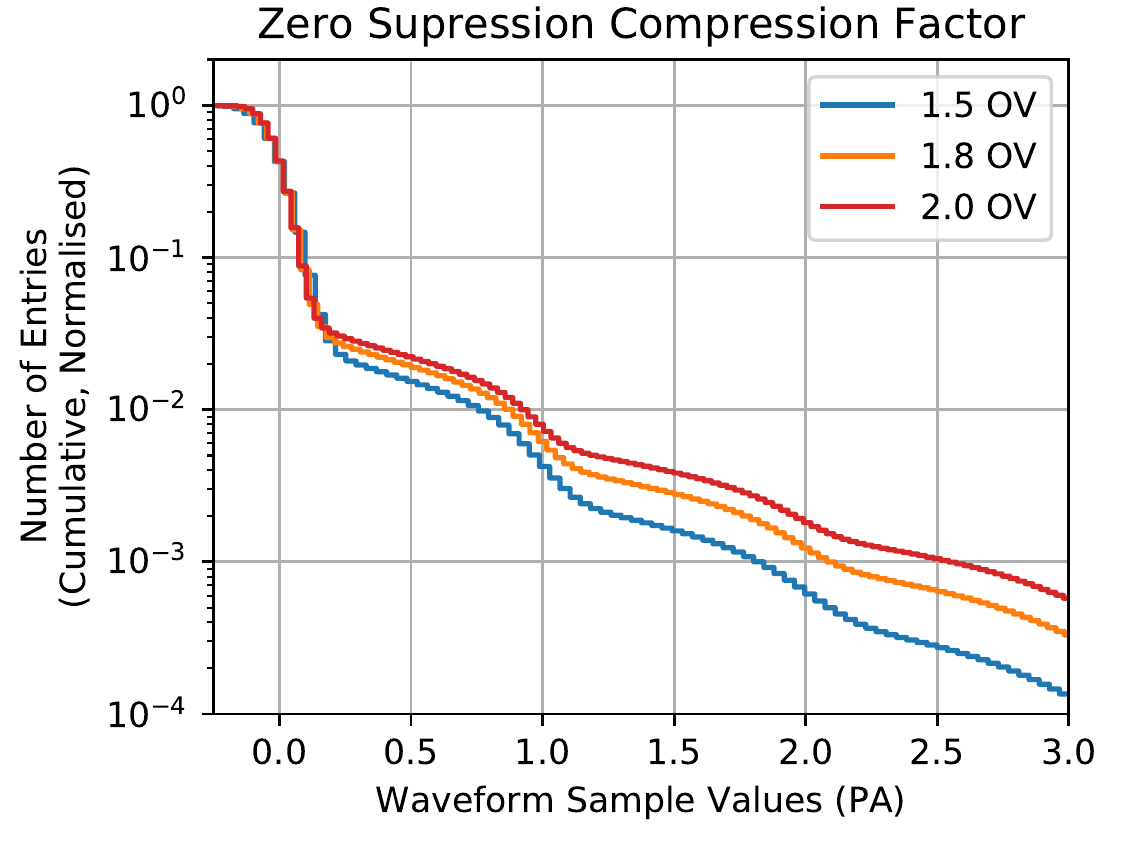}
\caption{The fraction of waveform samples above a given threshold. Zero suppression provides a significant waveform compression factor. The additional overhead for encoding the resulting gaps in the data-steam is not included in this figure.}
\label{fig_wf_compression}
\end{center}
\end{figure}

The majority of waveform sample values are near the pedestal, and are thus free of any SiPM signals. The fraction of samples above an applied lower threshold is shown in figure~\ref{fig_wf_compression} for different over-voltages. A zero suppression (ZS) value at 0.5\,PA is found to remove the pedestal contribution, whilst retaining all SiPM signals, resulting in a waveform compression factor of around 50. Increasing the threshold further to 1.5\,PA can provide another order of magnitude of waveform compression, at the expense of removing the single PA signals. 

\section{Trigger Strategy and Implementation}
\label{sec:trigger}
The required IBD signal efficiency and data reduction factor can be achieved with a single level of triggering and event building, which can be implemented in the per-plane FPGAs. This section begins with a description of the different per-plane trigger algorithms, followed by a description of the options for readout and event building. The section ends describing the implementation in firmware, and presents a summary of default settings used during physics data taking mode. The trigger rates stated are the sum across all detector planes.

\subsection{Triggers}
\subsubsection{IBD (i.e Neutron)}

\begin{figure*}
\begin{center}
\includegraphics[width=1.0\textwidth]{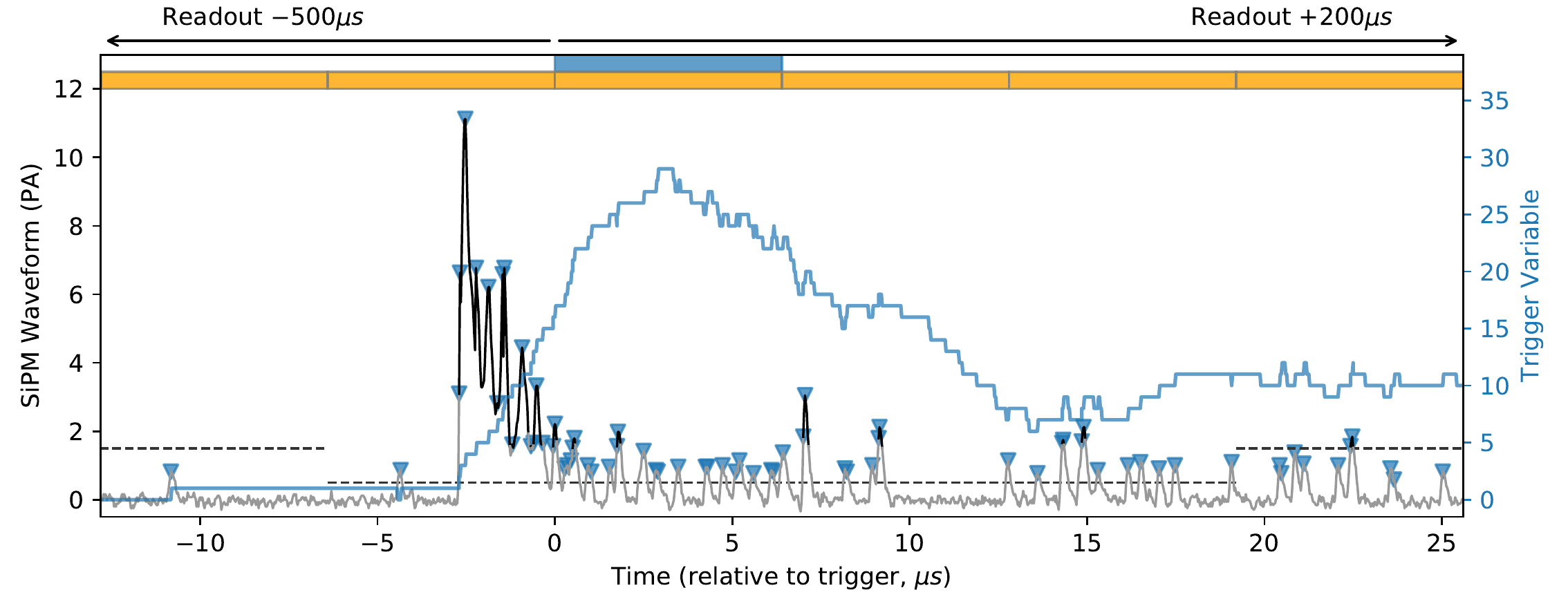}
\caption{Example neutron waveform (black). The dashed lines show the zero suppression threshold, which can be reduced local to the neutron trigger, as described in the text. The value of the neutron trigger variable (i.e number of peaks in the rolling time window), using algorithm parameters optimised for physics mode, is shown in blue. The rectangles at the top of the figure denotes the blocking of waveform samples, with the triggered block highlighted in blue. The arrows on the top indicate the time window that is read out before and after the trigger to include additional signals that can be used in offline analysis.}
\label{fig_neutron_diagram}
\end{center}
\end{figure*}

The sources and magnitudes of rates for various detector signals in physics mode are shown in table~\ref{tab:eventRates}. During reactor operation, the rate of gamma rays from the reactor that interact in the cube PVT, which can mimic a positron signal, is too high to allow efficient self-triggering on IBD positrons. Instead, the trigger strategy for IBDs relies solely on triggering on the IBD neutron. For each neutron trigger, a space-time region is read out, which is set large enough to encapsulate all signals from the IBD interaction. This allows the positron signals to be recorded without any trigger bias. The readout region recorded is considerably larger than the expected extent of the IBD interaction in both space and time. This allows the recording of very low amplitude associated signals, which can be used in offline analysis to aid discrimination of signal (e.g annihilation gammas) and background (e.g proton recoils). 

\begin{table}
\small

\begin{center}

    \begin{tabular}{cc}
        \toprule \midrule
        Signal & Detector Interaction Rate \\
        \midrule
        Dark Count (SiPM) & 100 MBq \\
        Reactor $\gamma$ & 100 kBq \\
        Cosmic Muons & 100 Bq \\
        Neutron & 10 Bq \\
        IBD & 0.01 Bq \\
        \midrule \bottomrule
        \rule{0pt}{0.1ex} 
    \end{tabular}
    \caption{Order of magnitude rates of different signals detectable by the whole SoLid detector.}
    \label{tab:eventRates} 
    \end{center}

\end{table}

This strategy is possible if neutron signals can be efficiently identified at a sustainable rate. The amplitude of neutron ZnS(Ag) signals is low and similar to that of both reactor gammas and SiPM dark counts. The scintillation signals themselves are sporadic, and spread over several microseconds - an example waveform is shown in figure~\ref{fig_neutron_diagram}. Several trigger algorithms, each suitable for FPGA implementation, have been compared~\cite{twepp}. The algorithm offering best discrimination involves tracking the time density of peaks in the waveforms. Specifically, for each detector channel, the number of waveform local maxima, whose amplitude are above a peak-finding threshold, is counted in a rolling time window. The algorithm triggers when the number of peaks in the window exceeds a threshold. The algorithm has the following parameters to be set during deployment:

\begin{itemize}
    \item $T$: amplitude requirement on waveform local maxima to be counted as a peak (0.5\,or 1.5\,PA). 
    \item $W$: the size of the rolling time window. The window width should be set at a scale near the characteristic decay time of the ZnS(Ag) (i.e a few microseconds). 
    \item $N_{\mathit{peaks}}$: the number of peaks required in the window to trigger.
\end{itemize}

\begin{figure}
\begin{center}
\includegraphics[width=0.6\textwidth]{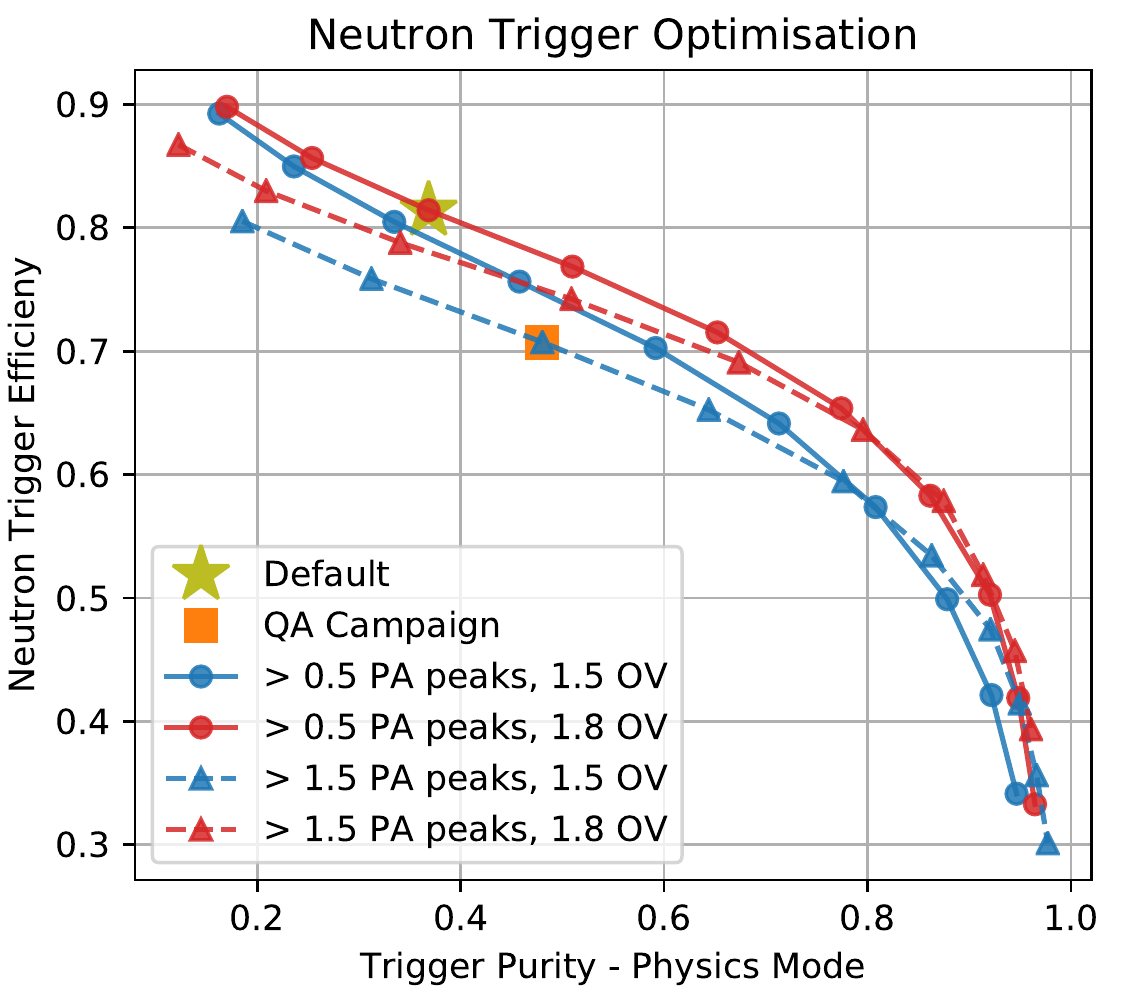}
\caption{ROC curves showing the optimisation of the neutron trigger. The contours show the dependence on both the SiPM over-voltage (OV) and the peak finding threshold $T_{\mathit{peaks}}$. Each point on a contour represents a number of peaks required to trigger in the rolling time window. The points take integer steps, where the $>$1.5 PA contours begin at 10 peaks, and the $>$0.5 PA contours begin at 16 peaks (both starting from the left). N.b the efficiency measurements were made using an AmBe source that was placed in a single position near the middle of the detector.}
\label{fig_trigTuning}
\end{center}
\end{figure}

\begin{figure}
\begin{center}
\includegraphics[width=0.6\textwidth]{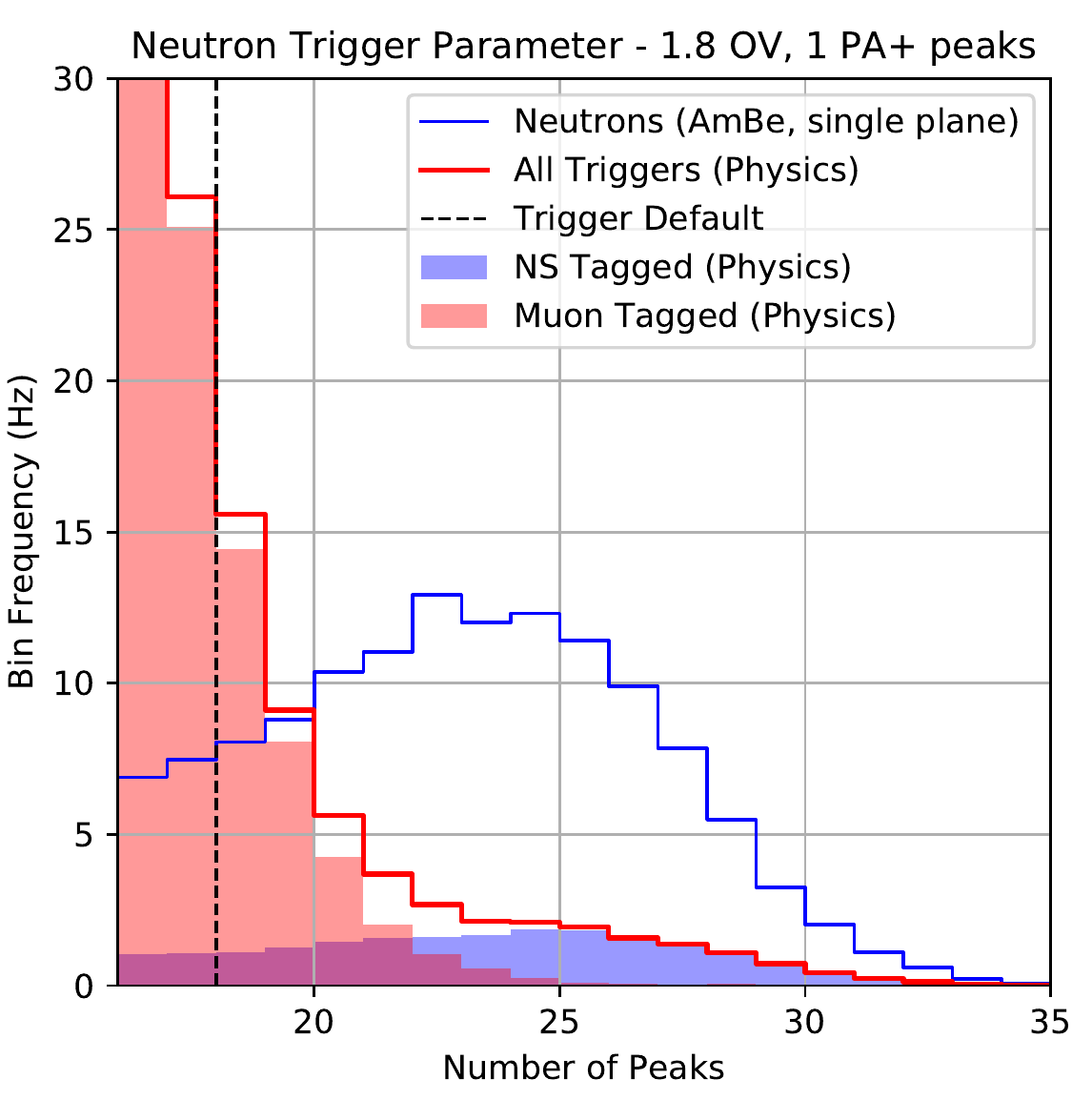}
\caption{Distributions of the trigger variable $N_{\mathit{peaks}}$ during physics mode. The equivalent for an AmBe source run, where the trigger is highly dominated by pure neutrons, is also included for reference.}
\label{fig_nvariable}
\end{center}
\end{figure}

$W$ is fixed at 256 waveform samples (6.4\,\textmu{}s). Scans of trigger efficiency vs trigger purity (i.e. receiver operating characteristic (ROC) curves) for different values of $N_{\mathit{peaks}}$ are shown in figure~\ref{fig_trigTuning}. An offline neutron ID algorithm, using pulse shape discrimination techniques, has been used to calculate the ROC curve values. This algorithm facilitates highly accurate tagging at negligible efficiency loss, with values of efficiency and purity that far exceed that of the trigger. Additional curves are included to scan both $T_{\mathit{peak}}$ and detector over-voltage. Optimal performance is achieved with the lower value of $T$ that includes counting 1 PA peaks, and with over-voltage at 1.8\,V. Further increases to the over-voltage result in unacceptable increases in SiPM pixel cross-talk. For $N_{\mathit{peaks}}$, the physics mode default values are taken from near the cusp of the optimal ROC curve, on the side of higher efficiency with a sacrifice in purity, corresponding to a trigger efficiency of 81\,$\pm$\,3\,\% (using measurements from a single position only). The neutron efficiency measurement arises from a comparison of the detected neutron rate with that expected from simulations, where the uncertainty is dominated by the systematic error assigned given the level of agreement between repeated measurements using multiple neutron sources. The corresponding purity is around 40\,\%. Further decreases in this threshold result in a too high data rate given long term data storage limitations. The distribution of the trigger variable $N_{\mathit{peaks}}$ for this configuration is shown in figure~\ref{fig_nvariable}. This study used neutrons tagged using the offline algorithm, for a data sample obtained using a less restrictive trigger threshold where $N_{\mathit{peaks}}$ $>$ 14. The muon identification used another offline ID based on topological information and energy deposits. A significant fraction of fake triggers are coincident with muon signals, particularly those muons whose incident angle is near the vertical or horizontal, and so deposit a large amount of energy along one particular fibre. Upon inspection of the waveform data, the most likely explanation is that these triggers arise from after-pulsing of the SiPMs due to the extremely large signals.

\begin{figure}
\begin{center}
\includegraphics[width=0.6\textwidth]{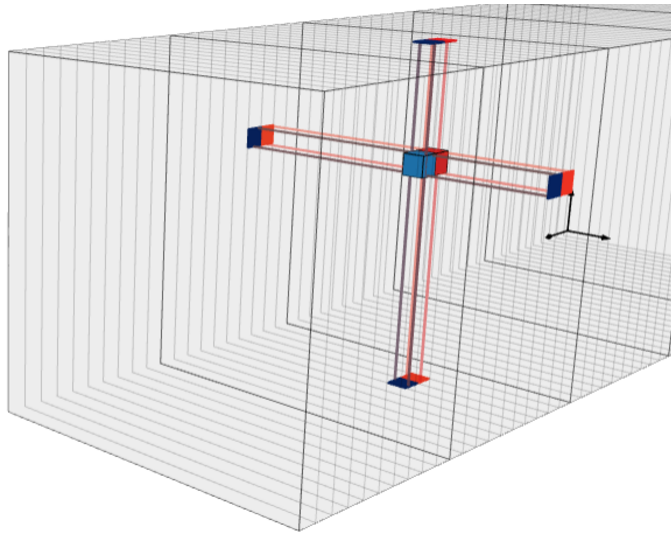}
\caption{Example EM (red) neutron (blue) coincidence candidate from the SoLid detector. The time separation between the signals is around 40\,\textmu{}s. All signals were read out using the neutron trigger only.}
\label{fig_ibd_eg}
\end{center}
\end{figure}

Simulations of neutrino IBDs and dominant backgrounds suggest that the majority of induced detector signals are contained within a few planes either side of the triggered detector plane. The default number of planes read out is three planes either side of the triggered plane. The neutron capture and thermalisation time follows an exponential distribution, with a mean lifetime of around 60\,\textmu{}s. However, some backgrounds have slower decay times around 120\,\textmu{}s. In order to study these backgrounds without bias from signal, a large time window of 500\,\textmu{}s before the trigger is read out. Finally, in order to study accidental backgrounds, a window of 200\,\textmu{}s after the trigger is also read out. The large time windows compared to the neutron thermalisation time also allow for isolation selections to be applied, which can be used for background tagging. An example IBD candidate event is shown in figure~\ref{fig_ibd_eg}. The corresponding data rate of the neutron trigger is 15\,MB/s for a trigger rate of around 40\,Hz, and does not change significantly depending on reactor operations.

\subsubsection{High Energy}
High energy signals, such as muons that enter the detector preceding IBD candidates, can be used to discriminate backgrounds from signal. An amplitude threshold trigger, targeted at PVT signals, has been implemented. An X-Y coincidence condition is also imposed to remove triggers caused by SiPM dark counts. The default physics mode threshold is 2\,MeV, which gives a trigger rate of 2.1\,kHz and data rate of 2\,MB/s during reactor-on periods. These rates decrease by around 10$\%$ during reactor-off periods. 

\subsubsection{Zero-bias Trigger}
A free running periodic trigger is used to allow continuous unbiased monitoring of the stability of the SiPMs, and any external noise contributions. The default trigger rate is 1.2\,Hz, and upon triggering, the entire detector is read out for a time window of 512 samples (non-zero-suppressed), giving a data rate of 3.9\,MB/s.

\subsection{Event Building and Zero Suppression}
\label{sec_event_building}
By default, all detector channels in the triggered detector plane are read out for each trigger. This is useful to reconstruct complicated event topologies offline, which include low amplitude signals that are shared between multiple fibres. A history time buffer is present for each channel, allowing data from both before and after the trigger to be read out. The exact width of the time window can be set for each trigger type, with limits depending on the buffer size and waveform compression rate due to zero suppression. The readout space region is set by propagating trigger signals to neighbouring detector planes, where the number of planes read out either side of the triggered plane depends on the trigger type. In cases of overlapping space-time regions due to triggers in close proximity, the readout time window is extended as required for each affected detector plane. 

All channels are zero-suppressed with a default threshold of 1.5\,PA in the absence of a trigger. The threshold can be lowered to 0.5\,PA for a short time region before and after a trigger for all plane channels depending on the trigger type. In cases of overlapping triggers, the lowest of the ZS thresholds is applied. This is especially useful for the neutron trigger, since a substantial fraction of the waveform is formed of very low amplitude signals, which can thus be retained for offline PSD. This is demonstrated in figure~\ref{fig_neutron_diagram}, which shows the regions where the ZS was reduced to 0.5\,PA from 1.5\,PA. 

\subsection{Firmware Implementation}
\begin{figure*}
\begin{center}
\includegraphics[width=1.0\textwidth]{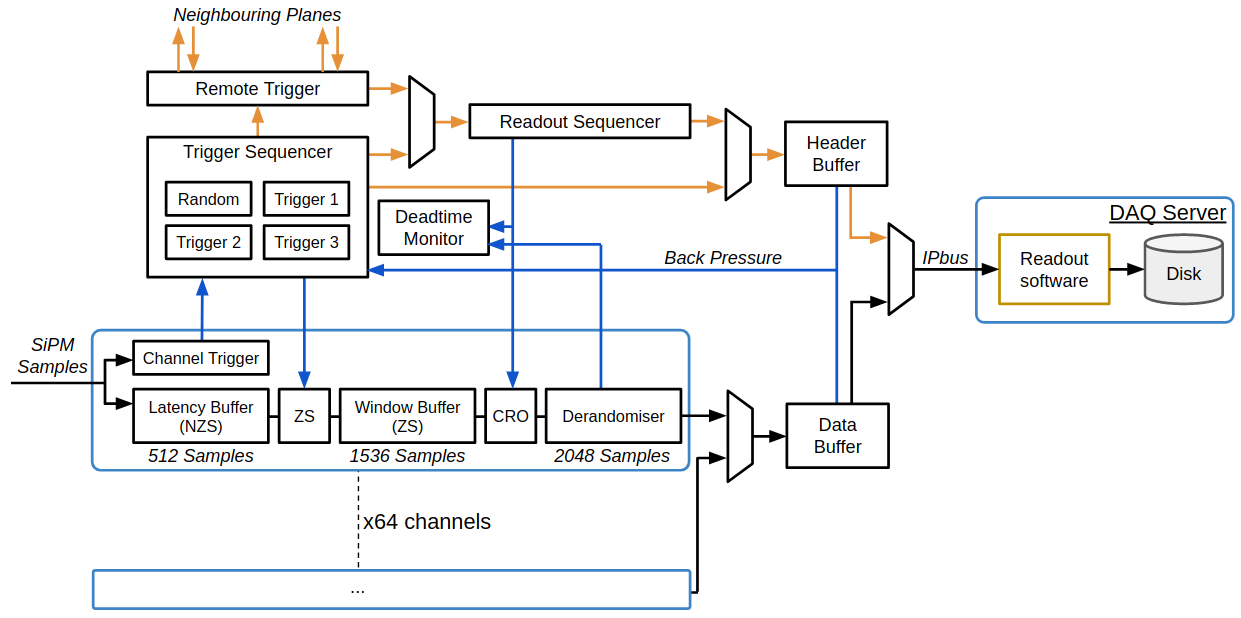}
\caption{Block diagram of the plane level FPGA firmware implementation.}
\label{fig_fm_diagram}
\end{center}
\end{figure*}

A block diagram of the per-plane FPGA firmware implementation is shown in Figure~\ref{fig_fm_diagram}. Data from the ADCs are deserialised and first placed in a FIFO latency buffer, which is non-zero-suppressed (NZS) and can store up to 512 waveform samples. This buffer is used to delay while a trigger decision is made. Alternative data sources are possible for debugging, such as a playback mode for pre-recorded data, as well as a signal generator capable of producing positron and neutron like signals. In parallel to filling the NZS buffer, the channel trigger operates for the threshold and neutron triggers, producing trigger primitives quantities on a sample-by-sample basis.

Each detector plane has a trigger sequencer that is responsible for the handling of triggers. In order to save FPGA resources, instead of triggering on a sample-by-sample basis, plane-level trigger decisions are made at a reduced rate based on the channel triggers primitives. A grouping of 256 sequential waveform samples is known as a \textit{`block'}, and trigger decisions operate on a block-by-block basis. The use of blocking has no effect on the physics, but substantially reduces the complexity and resource requirements of the firmware. A trigger occurs when a channel trigger fires in that block, or if the per-plane random trigger fired. Each trigger type can be configured to send a remote trigger to a pre-set number of detector planes either side of the trigger plane, up to the full detector. Given the long decay time of neutron signals, the neutron trigger often causes two sequential blocks to trigger. In these cases, the readout region is extended in time, and the trigger rates quoted previously are corrected for this double counting effect.

Data leaving the NZS buffer are zero-suppressed and enter the window buffer. A default ZS threshold is always applied, but the trigger sequencer can change the threshold depending on the trigger type. The duration stored in this buffer depends on its size and the ZS compression rate - a compression factor of 50, due to ZS at 0.5\,PA, allows up to 2\,ms to be stored. In practice, to avoid buffer overflow, the buffer is limited to 500\,\textmu{}s, which is suitably large for the IBD buffer for each neutron trigger.

Each plane has a readout sequencer that determines which data are recorded for each trigger (either local or remote triggers). This sequencer sets the duration of data readout from both before the trigger (i.e stored in the ZS buffer), and after the trigger (i.e not yet entered the ZS buffer) in units of blocks. The smallest amount of data that can be read out is therefore a single block, corresponding to 256 waveform samples (6.4\,\textmu{}s). Blocks are discarded or transferred to a per-channel derandomiser buffer depending on the decision of the readout sequencer. Finally, for each block, data from all channels are concatenated and stored in a data buffer for read out over IPbus by the local DAQ disk server. In cases where the derandomiser or data buffer overflows (e.g due to a high trigger rate), triggers are halted and the detector plane is in \textit{plane dead time}. Should the overflow occur during the concatenation of channel data, some channels may be excluded, incurring \textit{channel dead time}. Once the overflow has been cleared, triggers resume and the dead time periods are encoded in the data steam.

\begin{figure}
\begin{center}
\includegraphics[width=0.6\textwidth]{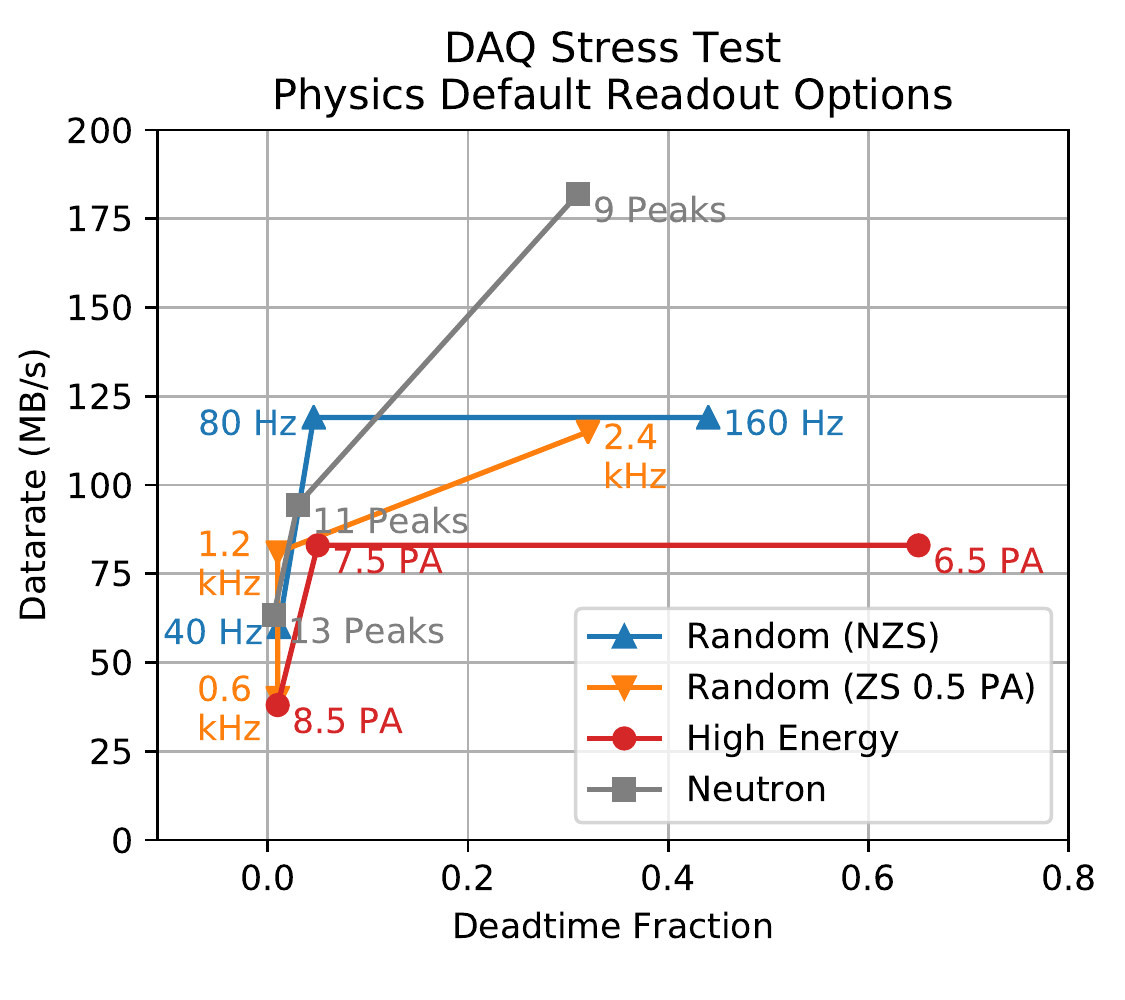}
\caption{Datarate v.s dead time fraction for varying trigger thresholds, for each physics mode trigger. The plane dead time dominates the channel dead time for each trigger configuration in all cases except for the NZS random trigger, where the channel dead time dominates. The plane dead time is taken as the fraction where the detector plane was in dead time during a full run, where as the channel dead time is taken as the fraction of channels returned compared to that expected.}
\label{fig_stress_test}
\end{center}
\end{figure}

\subsection{Trigger Settings Summary}
Table~\ref{tab_trig_summary} gives a summary of the different trigger settings presented. The mean recorded data rate of the experiment is 21\,MB/s during physics mode, which is dominated by the neutron trigger, and corresponds to around 1.6\,TB/day. A potential future upgrade of the experiment would be the addition of a software-level trigger, which could perform online event filtering and more sophisticated neutron ID techniques. This would increase flexibility during the optimisation of readout space-time regions and zero suppression thresholds, as well as providing further online data reduction. 

\section{System Performance}
\label{sec:performance}

\begin{table}
\scriptsize

\begin{center}
\begin{tabular}{ccccccc}
\toprule \midrule
\begin{tabular}[c]{@{}c@{}}Trigger \\ Type\end{tabular} & \begin{tabular}[c]{@{}c@{}}ZS \\ Threshold\end{tabular} & Condition & \multicolumn{2}{c}{\begin{tabular}[c]{@{}c@{}}Readout \\ Region\end{tabular}} & Rate (HZ) & Data rate (MB/s) \\
 &  &  & Space & Time (\textmu{}s) & & \\ 
 \midrule \midrule
Random & Disabled & Periodic 1\,Hz & Whole Detector & 12.8 & 1.2 & 3.9 (19$\%$)\\ \midrule
High Energy & 1.5\,PA & Waveform Sample $>$50\,PA & Triggered plane & 6.4 & 2.1k & 2 (10$\%$) \\ \midrule
Neutron & 0.5\,PA & \begin{tabular}[c]{@{}c@{}} $N_{\mathit{peaks}}$ $>$17 peaks \\  ($W_{\mathit{width}}$ = 6.4\,\textmu{}s, $T_{\mathit{peak}}$ $>$0.5\,PA)\end{tabular} & \begin{tabular}[c]{@{}c@{}}Triggered plane\\ $\pm$ 3 planes\end{tabular} & $-500$, $+200$ & 40 & 15 (71$\%$)\\
\midrule \bottomrule
\rule{0pt}{0.1ex} 
\end{tabular}
\caption{
Summary of trigger settings during reactor-on physics data taking. The only value that changes with the reactor power is the threshold trigger rate, which decreases by around 10$\%$ for reactor-off. The ZS threshold values are valid in the time window $\pm$ 2 blocks around the trigger - otherwise, such as the IBD buffer, the default ZS threshold of 1.5\,PA is applied.
}
\label{tab_trig_summary}
\end{center}
\end{table}

\subsection{DAQ Performance}
\label{sec:DAQ_performance}
The readout system has been stress tested to find the maximum data rate limit before incurring significant dead time. This has been performed for each trigger type separately by adjusting the respective trigger thresholds to increase data flow. The data rate as a function of dead time fraction is shown in figure~\ref{fig_stress_test}. The default readout options, as presented in the previous section, were used. In all trigger configurations tested, the dead time fractions are negligible below a data rate of 50\,MB/s. During default physics mode running, the combined effect of plane and channel dead time results in a data loss of $<$\,3\,\%. At higher data rates, the bottleneck currently arises in the readout software. Further optimisations are being investigated, the current rates are acceptable given the system requirements. 

\subsection{Run Control}
During physics mode, runs are taken sequentially and typically last 10 minutes in duration. This is set for convenience of offline file transfer - the size of the single file for a run is around 10\,GB. A hot start option is implemented for a quick restart between runs where no re-configuration is required - a firmware soft-reset is performed, which resets the data pipeline and empties the firmware buffers, and the next run begins. In this configuration, the up-time of the detector is around 95\%, with 5\% loss due to run hot starts. The data acquisition system is robust, with no major technical failures since physics running was established.  

The readout software runs on a disk server located next to the detector, which provides 50\,TB of local storage. This is split into two data partitions, which are periodically swapped and cleared. Data is first transferred to the Brussels VUB HEP Tier 2 data centre, and subsequently backed up at two further sites in France and the UK. GRID tools~\cite{dirac}, originally developed for LHC experiments, are utilised for data transfer and offline processing tasks. 

\subsection{Online Monitoring}
Run control operations are controlled via a dedicated python based web application, the \textit{SoLid Data Quality Monitor} (SDQM), which also provides an interface to inspect data quality. A small fraction of each run is processed online using the SoLid reconstruction and analysis program (Saffron2), and the output measurements and distributions can be inspected via the web application. Long term trends of certain measurements, such as SiPM gains and detector deadtime, are also stored in an online database, and input to an automated alarm system. 

In addition to detector measurements, several in situ environmental sensors are placed in the detector container and immediate vicinity, which are read out periodically via the SDQM and stored in the online database. 

\subsection{Detector Stability}
\begin{figure*}
\begin{center}
\includegraphics[width=1.0\textwidth]{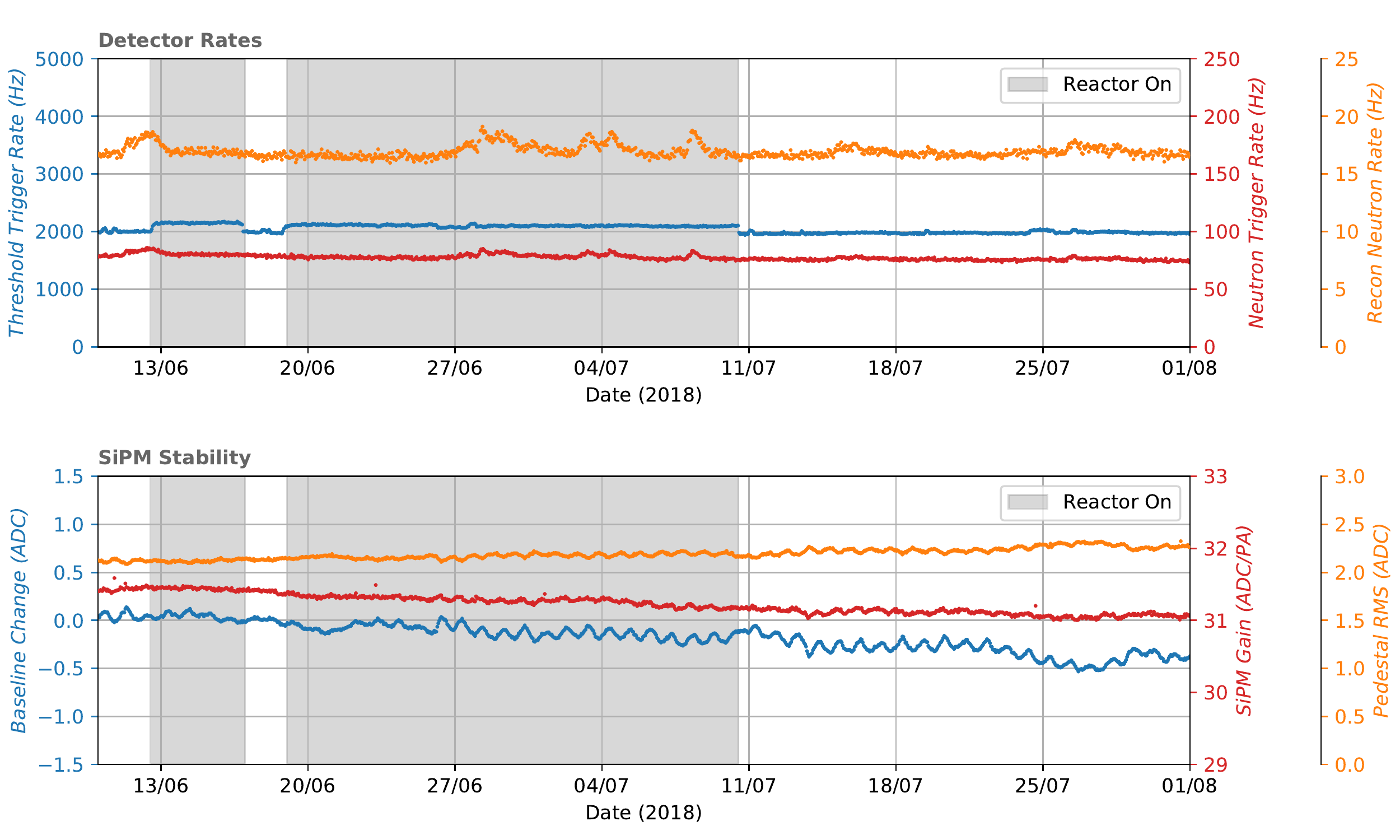}
\caption{Weekly trends of various detector metrics. The threshold trigger rate is the only metric sensitive to the reactor on-off transition.}
\label{fig_trends}
\end{center}
\end{figure*}

Figure~\ref{fig_trends} shows long term trends of both trigger rates and SiPM measurements. The transition between reactor-on and reactor-off can be seen in the small changes in the threshold trigger rate. Whilst the reactor is on, the standard deviation of the trigger rates for the neutron and threshold triggers, measured over a 1\,hr period, are 2\% and 1\% respectively. 

The SiPM measurements show small changes over the long term, as well as day-night variations. These changes are correlated with temperature changes up to 0.5$^{\circ}$C inside the detector container - the small increase in temperature increases the dark count rate of the SiPMs, causing the pedestal to slightly decrease, although the trigger rates remain unaffected. 

\section{Conclusion}
The SoLid detector readout system was successfully commissioned in late 2017. Custom electronics have been designed, constructed, checked for quality assurance, and deployed at the experiment site. Multiple triggers and data reduction techniques have been implemented at the FPGA level, including a novel PSD algorithm to trigger on neutron-like signals, with a trigger efficiency around 80$\%$ after optimisation. The corresponding data reduction factor is 10$^4$, and includes storing SiPM waveforms for offline analysis. Flexible space-time regions can be readout for each trigger type. Of the 3200 detector channels, 99$\%$ are operational, with a pedestal rms of 2 ADC (6$\%$ compared to a gain of 32 ADC/PA). The detector response is highly uniform across all detector channels. Physics data taking was established in early 2018. The total cost of the system is \$220k, or \$70 per channel. The system is highly stable, with an up-time of 95\% during physics running. 

\section{Acknowledgements}
This work was supported by the following funding agencies: Agence Nationale de la Recherche grant ANR-16CE31001803, Institut Carnot Mines, CNRS/IN2P3 et Region Pays de Loire, France; FWO-Vlaanderen and the Vlaamse Herculesstichting, Belgium; 
The U.K. groups acknowledge the support of the Science \& Technology Facilities Council (STFC), United Kingdom; We are grateful for the early support given by the sub-department of Particle Physics at Oxford and High Energy Physics at Imperial College London. 
We thank also our colleagues, the administrative and technical staffs of the SCK$\bullet$CEN for their invaluable support for this project. Individuals have received support from the FWO-Vlaanderen and the Belgian Federal Science Policy Office (BelSpo) under the IUAP network programme; The STFC Rutherford Fellowship program and the European Research Council under the European Union's Horizon 2020 Programme (H2020-CoG)/ERC Grant
Agreement n. 682474 (A. Vacheret); Merton College Oxford.

\end{document}